\documentclass[aps,nofootinbib,amssymb,amsmath,floats,showpacs,showkeys]{revtex4}
\newcommand{\bastar}{\begin{eqnarray*}}
\newcommand{\eastar}{\end{eqnarray*}}
\newskip\humongous \humongous=0pt plus 1000pt minus 1000pt

\newif\ifdtup

\relax
\newcommand{\be}{\begin{equation}}
\newcommand{\ee}{\end{equation}}
\newcommand{\ba}{\begin{array}}
\newcommand{\ea}{\end{array}}
\newcommand{\bea}{\begin{eqnarray}}
\newcommand{\eea}{\end{eqnarray}}
\newcommand{\pro}{\partial}
\newcommand{\n}{\hat n}
\newcommand{\oneg}{\difrac{1}{g}}
\newcommand{\valpha}{\vec \alpha}

\newcommand{\vX}{{\vec X}}
\newcommand{\vA}{{\vec A}}
\newcommand{\difrac}{\displaystyle\frac}
\newcommand{\nn}{\nonumber}

\newcommand{\balpha}{{\bar \alpha}}
\newcommand{\bbeta}{{\bar \beta}}
\newcommand{\bgamma}{{\bar \gamma}}

\newcommand{\bsigma}{{\bar \sigma}}

\begin{document}
\title{Torsion as a dynamic degree of freedom of quantum gravity}
\bigskip
\author{Sang-Woo Kim}
\affiliation{School of Physics, College of Natural Sciences, \\
Seoul National University, Seoul 151-742, Korea}
\author{D. G. Pak}
\email{dmipak@phya.snu.ac.kr}
\affiliation{ Center for Theoretical Physics, Seoul National
University,
Seoul 151-742, Korea}
\affiliation{Institute  of  Applied
Physics,  Uzbekistan  National  University,
Tashkent  700-095, Uzbekistan}
\begin{abstract}
The gauge approach to gravity  based on the local Lorentz group
with a general independent affine connection $A_{\mu cd}$ is
developed. We consider $SO(1,3)$ gauge theory with a Lagrangian
quadratic in curvature as a simple model of quantum gravity. The
torsion is proposed to represent a dynamic degree of freedom of
quantum gravity at scales above the Planckian energy. The
Einstein-Hilbert theory is induced as an effective theory due to
quantum corrections of torsion via generating a stable
gravito-magnetic condensate. We conjecture that torsion possesses
an intrinsic quantum nature and can be confined.
\end{abstract}

\maketitle

\section{Introduction}

The idea that torsion should play an important role
in gravity descends from works by E. Cartan \cite{cartan}
who had deep insight into the nature of space-time geometry.
Cartan was first to realize the tensorial character of torsion and
subsequently the approach to gravity including torsion was developed
extensively \cite{ivan1,hehl,odints,shapiro}.
Near that time H. Weyl invented a gauge invariance
principle \cite{weyl} which was
used successfully as a guiding rule
in constructing the modern theories
of electro-weak and strong interactions.
The gauge approach to gravity based on Lorentz and Poincare group
was proposed in \cite{uti,kibble,transl,ivan2} and later
was developed
by many physicists \cite{cho11,cho12,anttomb,carmeli1}.
The translational gauge formalism of Einstein theory was further developed in
\cite{cho12}.
In these approaches the initial classical Lagrangian
includes the Einstein-Hilbert term
(in addition to other possible terms quadratic in curvature and torsion)
providing the correct limit of Einstein gravity. Note
that in higher-derivative quantum gravity models with the
Einstein-Hilbert term the unitarity problem of the
physical $S$-matrix can be resolved consistently \cite{anttomb}.
The Carmeli  theory \cite{carmeli1}
based on the Lorentz gauge group  $SO(1,3)\simeq SL(2,C)$
is an example of a model which contains only a quadratic
curvature term of Maxwell type and, therefore, appears to be a fourth-order
theory for the metric tensor. For pure gravity
without matter the equations of motion in Carmeli's model lead to the
Newman-Penrose form of the vacuum Einstein-Hilbert equations.
This is an interesting hint
that the $SO(1,3)$ connection can be regarded as a dynamical
variable just like the metric tensor. Such a point of view was
adopted in \cite{martell} where authors treated the gauge
connection in Carmeli's model
as an independent variable and demonstrated the
renormalizability of the theory.

Einstein gravity is an effective theory
and can be deduced from some more fundamental theory
was discussed by Zel'dovich and Sakharov in 1967 \cite{zeld,sakhar}.
The possibility of inducing the Einstein theory
via quantum corrections was considered in past by many physicists
in various approaches: scale (conformal) invariance breaking schemes
\cite{adler1,adler2},
non-linear realizations of the Lorentz group
\cite{tseytlin,leclerc}, models with spontaneous symmetry
breaking \cite{ogiev,isham1}
where the graviton emerges as a Goldstone boson
\cite{zee,kirs,pirog}, and others \cite{macdow,pereira}.
In most of these approaches the Einstein-Hilbert term is induced by
quantum corrections due to interaction with a scalar matter field.
The disadvantage of this is that
we have no any experimental evidence for the existence
of fundamental scalar fields (like a dilaton or inflaton).
Moreover, the inducing of Einstein gravity from the interaction
with matter fields looks unsatisfactory because such a mechanism does not give
any answer to the origin of the classical Einstein theory
limit in the absence of matter.
Contrary to this, we use a minimal formulation of the theory
without scalar fields and demonstrate that even in a pure quantum
gravity model with torsion the Einstein-Hilbert action can be
induced due to quantum dynamics of the torsion
via formation of a non trivial vacuum with a gravito-magnetic
condensate.

In the present paper we are constrained to follow the gauge approach
to gravity based on the Lorentz gauge group.
We do not consider a more general case which includes the
Poincare gauge group, although such a formalism can be
developed as well. Our motivation to follow only the Lorentz gauge
group approach is twofold.
First of all, the local Lorentz symmetry has its own deep physical meaning
since it reflects the presence of weak equivalence principle
which is a basic fundamental principle lying in the nature of the
gravitational force. We expect that contortion as a part
of Lorentz gauge connection could play a principal role
at quantum level. The second reason of our constrained consideration
is that we are looking for a possible mechanism of emergent Einstein gravity
which can be induced as an effective theory during phase transition
at Planckian scale. So that we are interested in such a theory where
the metric obtains its dynamical content after phase transition
at lower energy scale. The contortion could be a good candidate
to switch on the mechanism of dynamical symmetry breaking
through the quantum corrections.
We develop this idea by suggesting
that {\it the torsion (contortion) represents exactly
the dynamic variable of quantum gravity.
Moreover, we conjecture that
torsion can be confined and exists intrinsically as a quantum object,
and its quantum dynamics manifest themselves by
inducing the Einstein-Hilbert theory as an effective theory of quantum
gravity below the Planckian scale}.

The paper is organized as follows. In Section II we describe the
main lines of the model following Utiyama-Kibble-Sciama gauge
approach to gravity \cite{uti,kibble,carmeli1,martell} based on
the Lorentz gauge group. In Section III an effective quantum action
is calculated in one-loop approximation by integrating out
torsion degrees of freedom. It has been shown that the
corresponding effective potential possesses a non-trivial minimum
which provides a vacuum gravito-magnetic condensate. The
classical limit of Einstein-Hilbert theory is obtained as an
effective theory induced by the quantum dynamics of torsion.
In Section IV we outline some parallels between the quantum
gravity model with torsion and the quantum chromodynamics (QCD).
The last section discusses the non-positiveness of the classical Hamiltonian and
our main results.

\section{Utiyama-Kibble-Sciama gauge approach to gravity}

In early approaches to the formulation of gravity as a gauge
theory of the Poincare group \cite{uti, kibble,cho11,cho12} the
vielbein $e_a^\mu$ and Lorentz affine connection $A_\mu^{~cd}$
were introduced (we use Greek letters $\mu, \nu, ...$ for the
space-time indices and Latin letters $a,b,c, ...$ for the Lorentz
indices). For simplicity we will not consider the Poincare gauge
group and restrict ourselves with a case of the proper Lorentz
group $SO(1,3)$. The infinitesimal transformation of a Lorentz
vector $V_a$ is given by
\bea
&&\delta V_a= [{\bf \Lambda},V_a]=\Lambda_a^{\,b} V_b,
\eea
where ${\bf \Lambda} \equiv
\Lambda_{cd} \Omega^{cd}$ is a Lie algebra valued parameter, and
$\Omega^{cd}$ is a generator of the Lorentz Lie algebra. The
vielbein $e_a^\mu$ allows us to convert Lorentz indices into
space-time indices and vice versa. We assume that the vielbein is
invertible
\bea
&e_a^\mu e_{b\mu} = \eta_{ab} ,
\eea
and the signature of the flat metric $\eta_{ab}$ in the tangent
space-time is Minkowskian, $\eta_{ab}=diag(+---)$. Throughout the
paper we follow Kibble's approach \cite{kibble} and treat the
vielbein as a fixed background field which does not manifest any
dynamical content. This is not a principal limitation and one can
overcome it by generalization to Poincare group. The covariant
derivative with respect to the Lorentz group is defined in a
standard manner
\bea
D_a=e_a^\mu (\pro_\mu + g{\bf A}_\mu) ,
\eea
where ${\bf A}_\mu\equiv A_{\mu cd} \Omega^{cd}$ is a general
affine connection taking values in the Lorentz Lie algebra, and
$g$ is a new gravitational gauge coupling constant corresponding
to the Lorentz gauge group. For brevity of notation we will use a
redefined connection which absorbs the coupling constant.
The general gauge connection $A_{\mu cd}$ can be written
as the sum
\bea
&&A_{\mu c}^{~~d} = \varphi_{\mu c}^{~~d} (e) + K_{\mu c}^{~~d} , \label{split}
\eea
where $K_{\mu c}^{~~d}$ is a contortion and
$\varphi_{\mu c}^{~~d}(e) $ is a Levi-Civita spin connection
given in terms of the vielbein
\bea
\varphi_{\mu a}^{~~b}(e)&=&\difrac{1}{2} (e^{\nu b} \pro_\mu e_{\nu a}-
e_a^\nu e_\mu^c \pro^b
e_{\nu c}+\pro_a e_\mu^b
-e_a^\nu \pro_\mu e_\nu^b+e^{\nu b} e_\mu^c \pro_a e_{\nu c}-\pro^b e_{\mu a}).
\eea
The
original Lorentz gauge transformation has the form
\bea
\delta e_a^\mu &=& \Lambda_a^b e_b^\mu, \nn \\
 \delta {\boldsymbol \varphi}_\mu (e)&=& -\pro_\mu {\bf \Lambda}-
                 [{\boldsymbol \varphi}_\mu,{\bf \Lambda}], \nn \\
\delta {\bf K}_\mu &=& -[{\bf K}_\mu,{\bf \Lambda}], \label{eqI}
\eea
where ${\boldsymbol \varphi}_\mu \equiv \varphi_{\mu cd} \Omega^{cd}$,
and ${\bf K}_\mu \equiv K_{\mu cd} \Omega^{cd}$.
Let us consider the main lines
of Riemann-Cartan geometry (see e.g. \cite{hehl}).
The torsion and curvature tensors are defined in a standard way
\bea
&& [D_a,D_b]=T_{ab}^{\,\,c} D_c + {\bf R}_{ab} , \label{comm}
\eea
here, ${\bf R}_{ab} \equiv R_{ab cd} \Omega^{cd}$.
The non-holonomicity coefficients $C_{ab}^{\,\,c}$ are given by the equations
\bea
&& [\Delta_a, \Delta_b]=C_{ab}^{\,\,c} \Delta_c ,~~~~~~~~~~
 \Delta_a \equiv e_a^\mu \pro_\mu .
\eea

To make the derivative $D_\mu$ covariant
under the general coordinate
transformation one
includes the Riemann-Cartan connection $\Gamma_{\mu \nu}^\rho$ as well
\bea
&&  D_\mu V^\nu=\pro_\mu V^\nu +\Gamma_{\mu \rho}^\nu V^\rho.
\eea
The Riemann-Cartan connection is related to the general
Lorentz connection $A_{\mu c}^{~\,d}$ by definition, via
\bea
&& D_\mu e^{\nu a}= \pro_\mu e^{\nu a} +\Gamma_{\mu \rho}^\nu e^{\rho a}-e^{\nu b}
A_{\mu b}^{~\,a}=0. \label{covmetric}
\eea
The Christoffel symbol
$\hat \Gamma^\rho_{\mu \nu}=\hat \Gamma^\rho_{\nu \mu}$
is related to the Levi-Civita connection
by means of the reduced equation
\bea
&& \hat D_\mu e^{\rho a}= \pro_\mu e^{\rho a} +
\hat \Gamma_{\mu \rho}^i e^{\rho a}-e^{ib} \varphi_{\mu b}^{\,\,a} =0.
\eea
Solving this equation one can find a standard relationship between
$\hat \Gamma_{\mu \nu}^\rho$ and $\varphi_{\mu a}^{~~b}$
\bea
&&\hat \Gamma_{\mu \nu}^\rho=e^{\rho a}e_{\nu b}
\varphi_{\mu a}^{~~b}+e_{\nu b}\pro_\mu e^{\rho b}.
\eea
An antisymmetric part of the Riemann-Cartan connection defines the torsion
components in the holonomic basis
\bea
&&\Gamma^\rho_{\mu \nu}-\Gamma^\rho_{\nu \mu}=T^{~\rho}_{\nu\mu}.
\eea
The contortion components in the unholonomic basis
can be expressed in terms of torsion,
and conversely
\bea
&& T_{ab}^{\,\,c}=A_{ab}^{~~c}-A_{ba}^{~~c}+C_{ab}^{~c}=K_{ab}^{~c}-K_{ba}^{~c} ,\\
&& K_{abc} = \difrac{1}{2} (T_{abc}-T_{bca}+T_{cab}) .
\eea

Under decomposition (\ref{split})
the Riemann-Cartan curvature is split into two parts
\bea
&& R_{abcd}=\hat R_{abcd}+\tilde R_{abcd},  \\
&& \hat R_{abcd}=\hat D_{[\underline a} \varphi_{\underline b] \underline c}^{
~\,\underline d}+\varphi_{[a|c}^{~~\,\,e}\varphi_{b]e}^{~\,d}, \nn \\
&& \tilde R_{abcd}=\hat D_{[\underline a} K_{\underline b] \underline c}^{
~\,\underline d}+K_{[a|c}^{~~e} K_{b]e}^{~\,d} , \nn
\eea
where the underlined indices stand for indices over which the
covariantization has been performed, and we use a short notation
for the index antisymmetrization $[a,b] = ab-ba$.

One can define the classical Lagrangian for a pure
gravity with torsion which
contains only terms quadratic in curvature tensor
\bea
&&  {\cal L}=
-\difrac{1}{4} (\alpha R_{ab cd} R^{abcd}-4 \beta R_{ab} R^{ab}
 + \gamma R^2), \label{L0}
\eea
where $R_{ab} \equiv \eta^{cd} R_{acbd}$ and
$R\equiv \eta^{ab} R_{ab}$ and $\alpha, \beta, \gamma$ are real
numbers. The case  $\alpha=\beta=\gamma=1$ corresponding
to Gauss-Bonnet type Lagrangian is considered recently in \cite{pakGB}.
In the present paper we deal with Yang-Mills type
Lagrangian setting $\alpha=1, \beta=\gamma=0$.
The Yang-Mills type Lagrangian with the general Lorentz connection
constructed from $SL(2,C)$ dyad and vielbein was
considered by Carmeli \cite{carmeli1}.
It had been demonstrated that the corresponding equation
of motion after projection with the vielbein produces
the vacuum Einstein-Hilbert equation  in Newman-Penrose
form.
Later Martellini and Sodano considered Carmeli's model
treating the Lorentz gauge connection as independent of vielbein,
and proving its renormalizability \cite{martell}. The
Lagrangian (\ref{L0}) has been considered also in \cite{fair}
with studying the problem of the non-compactness of the
Lorentz group.

Since the Lorentz group is not compact
the Lagrangain ${\cal L}_0$ leads to a non-positive definite
Hamiltonian.
For that reason one usually
adds to the Lagrangian the Einstein-Hilbert term $\sqrt{-g} R$
which helps to resolve the non-unitarity problem connected with
negative energy states.
One should stress that we do not introduce
the Einstein-Hilbert term as in Utiyama-Kibble-Sciama
approach \cite{uti,kibble}.
Neither we introduce the $SL(2,C)$ spin frame
as Carmeli did in his approach \cite{carmeli1}
to make the correspondence
with Einstein-Hilbert vacuum
equations in the Newman-Penrose formalism.

There were some attempts to consistently quantize the
gauge theory with a non-compact structure group \cite{fair}.
An interesting example is given by the gauge theory for
a non compact Virasoro-Kac-Moody group
which can be quantized properly
due to spontaneous symmetry breaking \cite{choviras}.
We will show that our model implies the quantum
effective action with a non-zero
vacuum condensate which provides a new scale in the
theory. The presence of the new dynamically
generated scale can justify the self-consistency of
the quantization procedure
in our model with the initial non-positive definite
classical Hamiltonian.

We will adopt the point of view that even though
the classical action (\ref{L0}) does not
produce a positive definite
Hamiltonian, nevertheless,
a consistent quantum theory can be formulated.
Since the canonical quantization method fails to handle
the quantization problem we will apply the quantization
scheme based on the
functional integration in Euclidean space-time.
Within this quantization scheme the quantum
theory can be constructed since
the Lorentz group is locally isomorphic to the product
of compact unitary groups $SU(2) \times SU(2)'$ in Euclidean space-time.

\section{Effective action}

The general approach to derivation of a low energy effective theory
is based on integrating out all high energy (heavy massive)
modes in the generating functional of the effective
action while keeping light modes (massless or light particles) as a classical
background.
A simple example of such an effective theory is represented by the well-known
Euler-Heisenberg effective Lagrangian in quantum electrodynamics
(QED) \cite{euler,cho13} which includes quantum contributions
of electron loops while the massless photon is kept as a classical
external field.
In a similar way we will integrate out the contortion
(which is supposed to gain an effective mass dynamically)
keeping the massless gravitational field $e_\mu^a$ as a
fixed background.

In general, unlike the QED Euler-Heisenberg effective action,
the quantum corrections
can induce additional changes in the structure of the initial perturbative
vacuum and generate phase transitions to new phases
with non-trivial vacua. During the phase transition
some non-vanishing
vacuum field condensates may be formed. Such a phenomenon
occurs in QCD during the transition between deconfinement
and confinement phases where non-vanishing
quark and gluon condensates are generated. The presence of non-trivial
vacuum condensates leads to additional modification of
the classical Lagrangian and Green functions.

With this preliminary let us start
with a pure Yang-Mills type
classical action
\bea
 S_{cl}&=& \int d^4x \sqrt {-g}{\cal L}_0=-\difrac{1}{4} \int d^4x\sqrt {-g}
(\hat R_{\mu\nu cd}+\tilde R_{\mu\nu cd})^2 . \label{Lcl}
\eea
A special feature of this action is the presence of an additional
local symmetry under the following so-called quantum gauge transformations
\bea
&&\delta e_a^\mu = \delta {\boldsymbol \varphi}_\mu(e)=0, \nn \\
&& \delta {\bf K}_\mu= - \hat D_\mu \tilde {\bf \Lambda}-[{\bf K}_\mu,\tilde{\bf \Lambda}],
\label{eqII}
\eea
where $\tilde{\bf \Lambda}\equiv \tilde \Lambda_{cd} \Omega^{cd}$ and
the restricted covariant derivative
$\hat D_\mu =\pro_\mu+{\boldsymbol \varphi}_\mu$
is defined
by means of the Levi-Civita connection only.
The restricted derivative $\hat D_\mu$ is covariant only under
the original Lorentz gauge transformation (\ref{eqI}).
Notice, that the construction of the restricted derivative
has a deep analogy with the mathematical structure of Abelian
projection in quantum chromodynamics \cite{cho1}. We will consider
the physical implications of this analogy in
the next section.

To study the problem of whether vacuum condensates
can be formed one has to calculate first the effective action
$\Gamma[\tilde K; e]$ which is a generating functional
of one-point irreducible Green functions \cite{pesk}.
The "classical" field
$\tilde K_{\mu c}^{~~d}\equiv <K_{\mu c}^{~~d}>_J$
represents the vacuum averaged value of the field operator at the presence of
a source, and $e_\mu^a$  is a fixed background metric.
Then, from the structure of the
effective potential part of the effective action $\Gamma[\tilde K; e]$
one can find whether a non-trivial vacuum condensate can be generated.

To calculate the effective action one should split
the contortion $K_{\mu c}^{~~d}$
into the "classical" field $\tilde K_{\mu c}^{~~d}$
and the quantum fluctuating part $Q_{\mu c}^{~~d}$
\bea
&K_{\mu c}^{~~d} =\tilde K_{\mu c}^{~~d}+Q_{\mu c}^{~~d}.
\eea

Exact calculation of the effective action
for an arbitrary field $\tilde K_{\mu c}^{~~d}$ and given external
field $e_\mu^a$ is a hard unresolved problem.
To simplify the calculation we use
the important property of the Yang-Mills type Lagrangian (\ref{Lcl})
due to the additional symmetry under the quantum transformation.
Namely, the vacuum torsion condensate
$<0|\tilde R_{abcd}|0>$ can appear only in the form of a gauge covariant
additive combination $\hat R_{abcd}+<\!\tilde R_{abcd}\!>$
what follows directly from the general renormalization
properties of the model \cite{martell}. So that,
to find the functional dependence of the
effective potential $V_{eff} (\hat R+<\!\tilde R\!>)$
on $<\!\tilde R\!>$
we calculate first the effective potential $V_{eff} (\hat R)$
by setting $\tilde K_{\mu c}^{~~d}=0$, i.e., $K_{\mu cd}=Q_{\mu cd}$.
Then, after completing the calculation
we will restore the dependence on torsion condensate
$<\!\tilde R\!>$ by simple adding this term
to $\hat R$ in the final expression for
$V_{eff} (\hat R)$.
By this way the calculation
becomes technically much simpler and
very similar to derivation of the effective action
in $SU(2)$ gauge theory.

We follow the standard formalism
of quantization based on the functional integration \cite{pesk}.
The quantization procedure is performed with respect
to quantum gauge transformation (\ref{eqII}),
and the resulting effective action will keep the
original Lorentz gauge invariance (\ref{eqI}).
With the generalized Lorenz gauge fixing condition
$\hat D_\mu {\bf Q}^\mu=0$ one can find the
gauge fixing term ${\cal L}_{gf}$ and Faddeev-Popov ghost term
${\cal L}_{FP}$
\bea
&& {\cal L}_{gf} = \difrac{1}{2} tr (\hat D_\mu {\bf Q}^\mu)^2, ~~~~~~~~~~\\
&& {\cal L}_{FP} = tr \, (\bar {\bf c} \hat D^\mu(\hat D_\mu {\bf c})).
\eea
After taking integration by parts
the effective action can be written in the form
\bea
&&\exp \big(i\Gamma_{eff} \big) =
\int
{\cal D} {\bf Q}_\mu {\cal D} {\bf c} {\cal D} \bar {\bf c}
\exp \Big{\lbrace} \difrac{}{}
i\int d^4x\sqrt {-g} {\rm tr} \Big[ \difrac{1}{4}  \hat {\bf R}^2_{\mu \nu}
-\difrac{1}{2} {\bf Q}^{\mu} (g_{\mu \nu} \hat D \hat D -
4 \hat {\bf R}_{\mu \nu}) {\bf Q}^\nu
+\bar {\bf c} (\hat D \hat D) {\bf c}~\Big] \Big{\rbrace} .
\eea
The formal expression for the one-loop effective action is
simplified to
\bea
\Gamma_{eff}&=&S_{cl}-\difrac{i}{2}Tr \ln [(g_{\mu \nu}
(\hat D \hat D)_{ab}^{cd}
-2 \hat R_{\mu \nu}^{~~ef} (f_{ef})_{ab}^{cd})]
+ i Tr \ln [(\hat D \hat D)_{ab}^{cd}], \label{det1}
\eea
where $(f_{ef})_{ab}^{~cd}$ are the structure constants of the Lorentz Lie algebra.
It should be stressed that the background curvature tensor $\hat R_{\mu \nu cd}$
in the last equation represents an arbitrary non-constant background.
The rigorous way to calculate the full effective action
including both the real part and especially the imaginary part must
specify the field background. For a constant background
the calculation can be carried out in full analogy with the
case of $SU(2)$ QCD \cite{cho3,cho4}.

The functional determinants in Eq. (\ref{det1}) are not well-defined
in Minkowski space-time. As is known the adding
an infinitesimal number factor $-i \epsilon$ to the
bare Laplace operator in the full operator
$\hat D \hat D$ is conditioned by the requirement
of causality. This infinitesimal
addition defines uniquely the
Wick rotation to the
Euclidean space-time. In our case we should perform the
Wick rotation in the base space-time and in the
tangent space-time as well, so that the Lorentz group
in the Euclidean sector turns into the compact orthogonal
group $SO(4)\simeq SU(2)\times SU(2)'$.
With this the functional integral becomes well-defined.
Certainly there remains the problem of analytical continuation
of the final expressions from Euclidean space-time back to
Minkowski space-time.

It is convenient to use the Weyl representation
for Dirac matrices $\gamma^\mu$ in Van der Warden notation.
In Euclidean space-time the Weyl representation
is defined as follows
\bea
&& \gamma^\mu =  \left(\ba{c} 0 ~~~~\sigma^\mu_{\alpha \balpha} \\
       \bsigma^{\mu \balpha \alpha} ~ 0 \ea\right) , ~~~~~
 \sigma^\mu_{\alpha \balpha} = (i {\bf 1} , -\vec \tau), ~~~
 \bsigma^{\mu \balpha \alpha} = (i {\bf 1} , \vec \tau),
\eea
where $\vec \tau$ are Pauli matrices.
The Cartan algebra of the Dirac matrices is given by
\bea
\{ \gamma^\mu, \gamma^\nu \} = -2 \delta^{\mu\nu}.
\eea

With these notations one can easily convert any real antisymmetric Lorentz
tensor $V_{cd}$
into the complex symmetric $SU(2)$ spin tensor $V_{\alpha \beta}$
\bea
&&V_{\alpha \beta} = V_{cd} \sigma^{cd}_{\alpha\beta}, ~~~~~~~~~~~~~~~~~
\bar V_{\balpha \bbeta} = V_{cd}\bsigma^{cd}_{\balpha \bbeta}, \\
&&\sigma^{cd}_{\alpha \beta}= \difrac{1}{4} \epsilon_{\beta \gamma}
     (\sigma^c \bsigma^d - \sigma^d \bsigma^c )_\alpha^{~\gamma},~~~~~
 \bsigma^{cd}_{\balpha \bbeta}= \difrac{1}{4}\epsilon_{\balpha \bgamma}
(\bsigma^c \sigma^d - \bsigma^d\sigma^c )^\bgamma_{~\bbeta}. \nn
\eea
The spinor indices can be lowered and raised
with the help of the spinor metric $\epsilon_{\alpha \beta}
(\epsilon_{\balpha \bbeta})$.
 The inverse relations for the spin tensor $V_{\alpha\beta}$
in terms of the antisymmetric tensor $V_{cd}$ includes a dual tensor
$V^{*cd}=\difrac{1}{2}\epsilon^{abcd} V_{cd}$
\bea
\ba{ll}
& V^{\alpha\beta} \sigma^{cd}_{\alpha\beta}= V^{cd}+V^{*cd}, \\
& \bar V^{\balpha\bbeta} \bsigma^{cd}_{\balpha\bbeta}= V^{cd}-V^{*cd} .
\ea
\eea

In spinor notation the Lorentz algebra can be termed
with a complex generator $\omega_{\alpha \beta} = \difrac{1}{2}
           (\sigma_{cd})_{\alpha\beta} \Omega^{cd}$ (and its
complex conjugate $\bar \omega_{\balpha \bbeta}$)
\bea
 [\omega_{\alpha\beta},\omega_{\gamma\delta}] &=& \difrac{1}{8}
 (\epsilon_{\alpha\gamma} \omega_{\beta\delta}
+\epsilon_{\beta\gamma}\omega_{\alpha \delta}
+\epsilon_{\alpha\delta} \omega_{\beta\gamma}
 + \epsilon_{\beta\delta} \omega_{\alpha \gamma}).
\eea

Now we can express explicitly the local isomorphism
$SO(4)\simeq SU(2)\times SU(2)'$
by defining the generators of $ SU(2)\times SU(2)'$ Lie algebra
\be
\ba{ll}
& T^i \equiv -2 \vec \tau_\alpha^{~\beta} \omega_\beta^{~\alpha}, ~~~~~~~~~~\\
&  T'^i \equiv -2 \vec \tau_\balpha^{~\bbeta} \bar \omega_\bbeta^{~\balpha}.
\ea
\ee
Notice that the above relations allow us to convert any symmetric
second rank spin tensor $T_{\alpha\beta}$ into an
$SU(2)$ vector $T^i$, so that we can find the
$SU(2)\times SU(2)'$ Lie algebra generated by $T^i, T'^j$
\be
\ba{ll}
& [T^i, T^j] = i \epsilon^{ijk} T^k, \\
& [T'^i, T'^j] = i \epsilon^{ijk} T'^k, ~~~~~\epsilon^{123}=1.
\ea
\ee

With this one can write down any Lorentz Lie algebra valued object
in $SU(2)$ notation. For instance,  one has
the following $SU(2)\times SU(2)'$ decomposition
for the gauge
Lorentz parameter ${\bf \Lambda}$
\bea
\Lambda_{cd} \Omega^{cd}& =& \Lambda^{\alpha\beta} \Omega_{\alpha\beta}+
\Lambda'^{\balpha\bbeta}\Omega'_{\balpha\bbeta}
=-i(\Lambda^i T^i + \Lambda'^i T'^i).
\eea

In $SU(2)$ notation the gauge Lorentz transformation
of the affine connection $A_\mu^{cd} \Omega_{cd} =
A_\mu^i T^i + {A'}_\mu^i T'^i$ reveals a
"chiral" structure corresponding to
the group product $SU(2)\times SU(2)'$
\be
\ba{ll}
& \delta A_\mu^i T^i = -\pro_\mu \Lambda^i T^i -i [A_\mu, \Lambda], ~~~~~~~\\
& \delta {A'}_\mu^i T'^i=-\pro_\mu \Lambda'^i T'^i-i[A'_\mu, \Lambda'] .
\ea
\ee
We have a similar factorization property for the curvature tensor
\bea
&& R_{\mu \nu cd} \Omega^{cd} =-i( R_{\mu \nu}^i T^i+{R'}_{\mu \nu}^i T'^i).
\eea

The functional determinants in (\ref{det1})
are factorized into the direct product of $SU(2)$
determinants, and the effective action
takes a more simple form
\bea
\Gamma_{eff}&=& S_{cl}
-\difrac{i}{2} Tr \ln [(g_{\mu \nu} (\hat D \hat D)^{ij}-2\hat R_{\mu \nu}^k
        \epsilon^{kij})]
-\difrac{i}{2} Tr \ln [(g_{\mu \nu} (\hat D' \hat D')^{ij}-2 \hat R'^k_{\mu \nu}
 \epsilon^{kij})] \nn \\
&+& i Tr \ln [(\hat D \hat D)^{ij}]
+ i Tr \ln [(\hat D' \hat D')^{ij}] , \label{funcdet}
\eea
where all quantities corresponding to the group $SU(2)'$
are marked with an apostrophe.

Notice, the $SU(2)$ curvature squared term $(\hat R_{\mu \nu}^i)^2$
contains not only the Riemann tensor $\hat R_{\mu \nu cd}$ but
also its dual counterpart
${\hat R^*}{}^{\mu\nu cd}$
\bea
\ba{ll}
(\hat R_{\mu \nu}^i)^2 =\difrac{1}{8} (\hat R_{\mu \nu cd} \hat R^{\mu \nu cd} +
   \hat R_{\mu \nu cd} {\hat R^*}{}^{\mu \nu cd})
 \equiv  \difrac{1}{8} (\hat R^2 + \hat R \hat R^*) , \\
(\hat R'^i_{\mu \nu})^2 = \difrac{1}{8} (\hat {R}^2 - \hat {R}
 \hat R^*).
\ea
\eea

In electrodynamics and Abelianized QCD \cite{cho3}
we can define two main gauge invariant variables
corresponding to the magnetic $(B)$ and
electric $(E)$ fields
(in an appropriate Lorentz frame)
\be
\ba{ll}
& B=\difrac{1}{2} \sqrt{\sqrt{F^4+(F F^*){}^2}+F^2}, \\
& E=\difrac{1}{2} \sqrt{\sqrt{F^4+(F F^*){}^2}-F^2},
\ea
\ee
where $F_{\mu\nu}$ is the Abelian field strength.
For a pure magnetic background one has
$F F^*=0$ and $F^2=2 B^2$. In a similar way as in Maxwell
electrodynamics we will consider the constant gravito-magnetic field
without specifying an explicit structure of the
corresponding connection
(or torsion) assuming that the following conditions
are satisfied
\bea
 \hat R \hat R^*=0, ~~~~~~~\hat R^2>0.\label{eq:conds}
\eea
For such a pure gravito-magnetic background one has
$(\hat R_{\mu\nu}^i)^2=(\hat R'^i_{\mu\nu})^2\equiv 2 H^2>0$.
With this
the functional determinants in (\ref{funcdet}) can be simplified
to a scalar form \cite{cho3}
\bea
&\Gamma_{eff}= S_{cl}
-i Tr \ln [\tilde D \tilde D-2 H] -i Tr \ln [\tilde D \tilde D+2 H]
-i Tr \ln [\tilde D' \tilde D'-2 H] -i Tr \ln [\tilde D' \tilde D'+2 H],
 \label{scaldet}\\
&\tilde D_\mu \equiv \pro_\mu +ig \tilde A_{\mu}, \nn
\eea
where $ \tilde A_\mu $ ($ \tilde A'_\mu $) is an Abelian projected component
of $SU(2)$ ($SU(2)'$) gauge connection.
Applying Schwinger's proper time
method of calculation of the effective action \cite{schwinger1,cho3}
and the $\zeta$-function regularization
we obtain the one-loop effective Lagrangian
\bea
&& {\cal L}_{eff} =-\difrac{1}{2} H^2-\difrac{11g^2}{48\pi^2}
       H^2(\ln\difrac{gH}{\mu^2}-c)+\difrac{ig^2}{8\pi} H^2, \label{eq:Leff}\\
&&c = 1-\difrac{1}{2} -\difrac{24}{11} \zeta(-1,\difrac{3}{2})=1.29214...\,. \nn
\eea
Formally, the expression for the effective Lagrangian
is identical to that of $SU(2)$ QCD.
Notice, the logarithmic term corresponds to the
nonperturbative contribution of all higher order
one-loop Feynman diagrams.
Now, using the renormalizability property of our model we can restore
the torsion vacuum field in the equation (\ref{eq:Leff})
by substitution $H^2\rightarrow
(\hat R+<\!\tilde R\!>)^2/2$.
With this the real part of the
effective potential $V_{eff}\equiv -Re({\cal L}_{eff})$
takes a final form
\bea
&& V_{eff} =\difrac{1}{4} (\hat R+<\!\tilde R\!>)^2+\difrac{11g^2}{96\pi^2}
    (\hat R+<\!\tilde R\!>)^2(\ln\difrac{g\sqrt{(\hat R+<\!\tilde R\!>)^2/2}}
{\mu^2}-c).
\eea
Obviously, the real part of the effective potential
has a new non-trivial minimum
at $\hat R=0$ and with a non-zero vacuum torsion condensate
$<\!\tilde R^2_{abcd}\!>$.
Notice, that for the vacuum state corresponding to
that minimum we have still an imaginary part of the
effective Lagrangian
$ig^2 <\!\tilde R^2\!>/16 \pi$
which is nothing but the Nielsen-Olesen
imaginary part derived in $SU(2)$ QCD long time ago \cite{savv}.
The appearance of the imaginary part is a natural consequence
of our constant field approximation which implies that
the vacuum is not stable
and can be treated only as an approximation to a true vacuum with
a lower energy.

To make some estimations
let us consider for simplicity the real part of the effective potential
with the torsion condensate in a flat metric
background, $\hat R_{\mu\nu cd}=0$,
\bea
V_{non-ren} &=&\difrac{1}{2} \tilde H^2+\difrac{11g^2}{48\pi^2}
  \tilde H^2(\ln\difrac{g \tilde H}{\mu^2}-c), \\
\tilde H^2 &=& \difrac{1}{2} <\!\tilde R_{\mu\nu cd}^2\!>. \nn
\eea
One can renormalize the effective potential $V_{non-ren}$
imposing an appropriate normalization condition
\bea
&& \difrac{\pro^2 V_{non-ren}}{\pro \tilde H^2}
\Big |_{\tilde H=\bar \mu^2}=\difrac{g^2}{\bar g^2}.
\eea
The renormalized effective potential includes
the running coupling constant $\bar g(\bar \mu)$ which
depends on energy scale parameter $\bar \mu$
\bea
&& V_{ren} = \difrac{1}{2} \tilde H^2 + \difrac{11{\bar g}^2}{48\pi^2}
   \tilde H^2(\ln\difrac{\bar g \tilde H}{{\bar \mu}^2}-\difrac{3}{2}).
\eea
One can check that the effective potential satisfies the
renormalization group equation
with the same $\beta$-function as that in
a pure $SU(2)$ Yang-Mills theory.
The effective potential has a non-trivial minimum
$V_{min}$ and leads to a
gravito-magnetic condensate $<\!\tilde H\!>$
\bea
&&V_{min} = -\difrac{11 \bar g^2}{192 \pi^2} <\!\tilde H\!>^2,  \\
&& <\!\tilde H\!>= \difrac{\bar \mu^2}{\bar g}
 \exp [-\difrac{24 \pi^2}{11 \bar g^2}+1] .\label{condens}
\eea

The presence of the minimum of the effective potential
does not guarantee that the corresponding new vacuum
is stable. The stability of the vacuum condensate
even in a pure $SU(2)$ model of QCD
presents a long-standing problem, and resolving that problem
has passed through
several controversial results since the early
papers by Savvidy, Nielsen and Olesen
\cite{savv,cox,ambj,sch,ditt}.
Without clear evidence or at least a strong indication
of vacuum stability one can not make any serious statement based
on the existence of a non-trivial vacuum condensate.
Recently, substantial progress in resolving this problem
in favour of stability of the magnetic vacuum
has been achieved \cite{cho3,cho4}.
Moreover, it has been found
 that a stable classical configuration
made of monopole-antimonopole strings does exist in the $SU(2)$
model of QCD \cite{ff}, providing a strong argument that a stable
magnetic vacuum can exist in QCD and, therefore, in
our gauge model of quantum gravity with torsion as well.

The gravito-magnetic condensate
$<\!\tilde H\!>$ generates a new renorminvariant scale
$M$ in the theory.
We suppose that the local gauge Lorentz symmetry
is not broken, so the lowest torsion condensate must vanish
$<T_{abc}>=0$.
Since one has a non-trivial dynamically generated scale
$M$
one can expect a non-vanishing vacuum averaged value for
the curvature tensor corresponding to the torsion
\bea
<\tilde R_{abcd}> = \difrac{1}{2} M^2 (\eta_{ac} \eta_{bd}-\eta_{ad}\eta_{bc}),
\label{assumpn}
\eea
The factor $M^2$ need to be positive since it
corresponds to a positive curvature space-time which can only be
created due to quantum fluctuations of torsion through the
vacuum transition from the trivial vacuum to the non-trivial one.
It is important to stress that the above expression for
$<\tilde R_{abcd}>$ with a positive scale $M^2$ satisfies
the conditions (\ref{eq:conds}) for the gravito-magnetic field.
Moreover, the tensor structure of (\ref{assumpn}) is unique
because adding another term proportional
to the antisymmetric tensor $\epsilon_{abcd}$ is forbidden
by (\ref{eq:conds}), and
such a term would contain a gravito-electric component
implying quantum instability.

Expanding the original classical Lagrangian (\ref{Lcl}) around the new
vacuum by shifting $\tilde R_{abcd} \rightarrow \tilde R_{abcd}
 + <\tilde R_{abcd}>$ one obtains
\bea
&&{\cal L}_{EHeff}  = -\difrac{1}{4} (\hat R_{abcd}+\tilde R_{abcd})^2
 =-\difrac{1}{4} \hat R^2 - \difrac{1}{2} \hat R M^2
-\difrac{3}{2} M^4. \label{EHeff}
\eea
One should emphasize that even though the vacuum averaged value
$<\tilde R_{abcd}> $ is specified by (\ref{assumpn}), the classical
gravitational field $\hat R_{abcd}$ is not constrained in general.
The last term in the equation corresponds to
a positive vacuum energy density which is supposed to
be born during the vacuum transition. From this we can estimate
the numeric value of $M$ by setting $3 M^4/2=|V_{min}|$
\bea
&& M^2 =\difrac{\bar g}{\pi}\sqrt{\difrac{11}{288}} <\!\tilde H\!>
=\difrac{\bar \mu^2}{\pi}
    \sqrt{\difrac{11}{288}} \exp [-\difrac{24 \pi^2}{11 \bar g^2}+1] .\label{scale}
\eea
Notice, that the scale $M$ is renorminvariant, i.e., it does not depend on
a particular value of $\bar \mu$.
The last two terms in (\ref{EHeff}) produce
the Einstein-Hilbert type terms in the effective Lagrangain
(in units $\hbar = c=1$)
\bea
{\cal L}_{EHeff}=- \difrac{1}{4} \hat R_{abcd}^2 - \difrac{1}{16 \pi G} (\hat R
+2 \lambda) .
\eea
So that the Newton constant $G$ and the cosmological constant $\lambda$
are determined by only one renorminvariant scale $M^2$,
in other words, by the renormalized running coupling constant $\bar g$
at some scale $\bar \mu$, expected to be of the order of the Planckian energy
$10^{19} Gev$.

Certainly, the assumption (\ref{assumpn})
leads to a desired
induced Einstein-Hilbert term, as has been known before.
The most important point is to provide a foundation
for that hypothesis. In our approach we put this assumption
on real ground by explicitly calculating the effective potential
and having found a stable classical vacuum solution
in $SU(2)$ gauge model \cite{ff} which is a part of
Lorentz gauge theory in Euclidean formulation.
We have hope that this assumption might be true, at least in the
framework of our simple model.

One can not fit both experimental values
of constants $G$ and $\lambda$ concurrently by
adjusting the scale parameter $M$,
so that the cosmological problem
remains unresolved.
An additional uncertainty is related with the unknown value of the
energy scale $\bar \mu$ which is only supposed to be of order Planckian
one, but for possible various phases the scale $\bar \mu$
may be different.
We consider two particular
cases when the scale parameter $M$ corresponds
to the experimental value of $G$ and $\lambda$.
In each case
we will find that the coupling constant
$\bar g$ takes large and small values respectively.
It is possible that
there are two phases corresponding to these strong
and weak coupling constants. Recently the existence of two
phases in gravity was suggested
in \cite{polonyi} within a different approach.

To justify the known value
for Newton's constant one should put the
following value
of the structure constant
$\alpha_g=\bar g^2/4 \pi \simeq 1.52$, which corresponds to a
strong coupling phase. Notice that
the cosmological constant is of the Planckian order.
This value is consistent with cosmological models which
elaborate the large value of the cosmological constant at
very early universe to provide
the fast initial inflation.

The weak coupling phase can be determined by the
known experimental bound for the vacuum energy density
\bea
&& \rho_v=\difrac{\lambda}{8 \pi G} \simeq 2\cdot 10^{-47} (Gev)^4.
\eea
This implies the following value of the scale
$M$ according to (\ref{EHeff}, \ref{scale})
\bea
M^2= 3.64 \times 10^{-24} (Gev)^2.
\eea
The corresponding value for the structure constant is small
\bea
&& \alpha_g=\difrac{\bar g^2}{4 \pi} =0.0123.
\eea
This value can be compared with the value  $\alpha_{SSGUT}\simeq 0.04$
of the structure constant
in supersymmetric $SO(10)$ GUT model at unification scale
$2 \times 10^{16} Gev$.
The same order of the structure constants $\alpha_g$ and $\alpha_{SSGUT}$
might be a hint to the origin of supersymmetry
and its relation to quantum gravity.
It would be natural to suspect such a connection
since the algebra of supersymmetry
had been invented exactly as an extension
of the Poincare Lie algebra which describes the
space-time symmetries intimately related to gravity.
The weak coupling phase can be considered as an analog
to quark-gluon plasma phase of QCD where some
intriguing results have been obtained recently \cite{plasma}.
Namely, this phase is described by the liquid model with
unexpected still non-vanishing gluon condensate.
It would be of interest to study the possible relationship
of the gravitational weak phase to the hydrodynamic modelling
the quantum gravity phenomenology proposed in \cite{hydro}.

\section{Parallels to QCD: torsion to be confined}

The consistent quantum theory of strong interaction is presented
by the quantum chromodynamics. One should stress that
at classical level the QCD can not serve merely as a classical
theory because single quarks and gluons
are not observable in principle as free particles.
The classical theory of strong interaction below the
confinement scale is described by other phenomenological
models where classical states are represented by hadrons, the bound states
of quarks and gluons.
There is a deep analogy
between our gauge model of gravity
with torsion and the theory of strong interaction
in mathematical structure.
It is a very intriguing question up to which extent
this analogy lies on.
We recall the main construction
of the gauge invariant Abelian projection
in $SU(2)$ QCD \cite{cho1}.
The Abelian decomposition of the full gauge potential
of $SU(2)$ gauge theory
has been implemented with a scalar triplet $\n$
which represents pure topological degrees of freedom
classified by the homotopy group $\pi_3(S^2) \simeq {\mathbb Z}$
\bea
 & \vec{A}_\mu =A_\mu \n - \oneg \n\times\pro_\mu\n+\vX_\mu
         = \hat A_\mu + \vX_\mu, ~~~~~
  (\n^2 =1,~~~ \hat{n}\cdot\vec{X}_\mu=0), \label{chodecomp}
\eea
where $\hat A_\mu$ is a restricted potential,
and $\vX_\mu$ represents the off-diagonal component,
the so-called valence gluon.
The important property of the restricted
potential $\hat A_\mu$ is that it possesses
the transformation properties of the full $SU(2)$ gauge
connection even in the absence of the valence gluon.
The scalar $\hat n$ is covariant constant
\bea
\hat D_\mu \n = \pro_\mu \n + g {\hat A}_\mu \times \n = 0
\label{constcov}
\eea
and has a clear mathematical origin
as an isometry vector corresponding to the
Cartan subalgebra $U(1)$ of
$SU(2)$ Lie algebra.
Under the infinitesimal gauge transformation
\be
\ba{ll}
&\delta \n = - \vec \alpha \times \n,\\
&\delta \vA_\mu =\oneg (\pro_\mu \vec \alpha+ g {\vec A}_\mu \times \vec \alpha)
    \equiv \oneg \vec D_\mu \vec \alpha
\ea \label{eqclassic}
\ee
one has the following transformation rules for the
Abelian and off-diagonal gluon components
\be
\ba{ll}
&\delta A_\mu = \oneg \n \cdot \pro_\mu \valpha , ~~~~~\\
&\delta \vX_\mu = - \valpha \times \vX_\mu  .
\ea
\ee
Notice, that the valence gluon
$\vec X_\mu$ transforms covariantly
like a vector and plays the role of charged matter.
In addition to the initial classical transformation (\ref{eqclassic}),
there is a second type of transformation
which originates from the additive structure
of the Abelian decomposition (\ref{chodecomp})
\be
\ba{ll}
& \delta \hat n  = 0 ,~~~~~\\
& \delta A_\mu = \difrac{1}{g} \hat n \cdot \vec D_\mu \valpha ,~~~~~\\
& \delta X_\mu= \difrac{1}{g} (\vec D_\mu \valpha -
        (\hat n \cdot \vec D_\mu \valpha) \hat n).
\ea
\ee

From the comparison of the $SU(2)$ QCD structure
with the Lorentz gauge model introduced in Section 2
one can immediately find the
analogy between the restricted potential $\hat A_\mu$ and valence
gluon $\vec X_\mu$ in QCD on the one hand,
and the Levi-Civita connection $\omega_{\mu cd}$
and contortion $K_{\mu cd}$ in Lorentz gauge model
on the other.

Obviously, in QCD (as well as in the Weinberg-Salam model which contains
the group $SU(2)$ as a subgroup) we can not treat the off-diagonal
component $\vec X_\mu$ as a true covariant vector. The reason is that
if we introduce, for instance, a mass term for the off-diagonal gluon
into the Lagrangian the renormalizability will be lost.
For the same reason we can not treat the contortion $K_{\mu cd}$
as a true tensor in gravity models in attempts
to formulate a quantum renormalizable theory.

Let us consider the following aspect of the
confinement problem in QCD regarding the
fact that quarks and gluons are not observable single
particles. One heuristic argument why
we can not observe the color single states is
the following \footnote[1]{ one of authors (DGP)
acknowledges Y.M. Cho for elucidating this argument.}:
quarks and gluons are not gauge
invariant and we have no a conserved color charge
like the electric charge in Maxwell theory.
So that quarks and gluons can not be observable
as single physical particles unless the color $SU(3)$ symmetry
breaks down. The reason why
we can observe the vector bosons in Weinberg-Salam
model is due to the spontaneous breaking of the symmetry.
Besides, in Maxwell theory the classical electric and magnetic fields
are gauge invariant concepts.
This is not the case with the non-Abelian theory of QCD
where gluons are not gauge invariant objects.
This fact along with the color symmetry being unbroken
gives a natural explanation to the confinement phenomenon
from the symmetry point of view.

If we accept the hypothesis that a Lorentz gauge model of
gravity with torsion possesses two types of gauge symmetry (\ref{eqI},\ref{eqII})
then we will be forced to accept the confinement
of torsion unless the quantum Lorentz gauge symmetry
breaks down. The only classical objects which can be observed
are bound states of $K_{\mu cd}$ and torsion vacuum
condensates. If torsion is confined this could help
to circumvent difficulties related with
available experimental limits and some theoretical
severe restrictions on propagating torsion
\cite{bel,pei}.

\section{Discussion}

Let us return to the problem of positive definiteness of
the Hamiltonian in the classical theory corresponding to the
Lagrangian (\ref{L0}).
Note that the existence of the quantum theory
when the corresponding classical theory can not be defined consistently
is not a new situation in quantum field theory.
The classical field theory of the
Dirac electron is not well-defined due to the existence of
negative energy states (Dirac sea).
Upon postulating the anticommutative relations for the
creation and annihilation operators the
quantum theory of electron becomes consistent,
but still the classical energy remains positive undefinite.
The deep origin of that problem was studied
in \cite{schwinger2,pauli,feynman}, and
it is related to transformation properties of
some operators under the time reflection.
As it was pointed out in \cite{schwinger2},
there was a principal contradiction between
the facts that the Lagrangian is a scalar function
under the time reversal meanwhile the energy-momentum vector
$P_\nu=\int d \sigma^\mu T_{\mu\nu}$ is a pseudovector,
and the electric charge $Q=1/c \int d\sigma^\mu j_\mu$
and the mass term of the electron $M \bar \psi \psi$
are pseudoscalars.
In path integral approach these facts allow
the interpretation of the negative energy states
as positrons moving backward in time \cite{feynman}.
The quantization based on the functional integration
is more general since it allows the quantization of non-quadratic
in momentum Lagrangians and, as in the particular example
of quantum electrodynamics, the path integral sums up all paths
including those that correspond to time reversed direction.
The compact group structure of the Lorentz group
in Euclidean space-time guarantees the consistent
quantization of the theory (in the present paper
we do not consider the problems
of analytical continuation from Euclidean space-time
back to Minkowski one).
Besides this we give a heuristic argument on a
possible resolution of the negative energy
problem of the classical Hamiltonian.
Let us consider the time reflection
operation $t \rightarrow -t$.
To find the properties of the contortion under the time reflection
it is convenient to
write down the linearized equations of motion
in flat space-time background in the generalized
Lorenz gauge $\pro_\mu K^{\mu cd} =0$
\bea
&& \pro^\mu \pro_\mu K_{\nu cd} = \bar \Psi (\gamma_\nu \Sigma_{cd} +
\Sigma_{cd}\gamma_\nu) \Psi.
\eea
Under time reversal the transformation rule for
a spinor is given by $\Psi \rightarrow \hat T \Psi$ with
$\hat T = \gamma_1 \gamma_2 \gamma_3$. This implies
$\bar \Psi \Psi \rightarrow  - \bar \Psi \Psi$,
i.e., the spinor matter density is a pseudoscalar.
From the equations of motion one can find the
transformation properties of the contortion under the time reflection
\bea
\ba{ll}
& K_{ijk} \rightarrow - K_{ijk},~~~~~~~~~K_{i0k} \rightarrow + K_{i0k} ,\\
& K_{0jk} \rightarrow + K_{0jk}, ~~~~~~~~~K_{00k} \rightarrow - K_{00k} ,
\ea
\eea
so that, $K_{\mu jk} $ manifests itself as a pseudovector
(like a photon $A_\mu$ or a pseudovector combination $\bar \psi \gamma_\mu \psi$
made of the Dirac spinor)
whereas the part of contortion $K_{\mu 0k}$, which gives a negative
energy contribution, represents the opposite properties of a vector.
Since the contortion is invariant under charge conjugation
it represents a neutral particle.
By analogy with
the path integral formulation of quantum mechanics
we can interprete $K_{\mu 0k}$
as a particle moving back in time with a
positive energy like
a positron in Feynman theory of positron \cite{feynman}.
Obviously such an analysis can not be performed
in a case of gauge models with a non-compact internal group.

In conclusion, we propose a simple gauge model with a local Lorentz group
which is supposed to describe the quantum theory of gravity with torsion.
One-loop effective action is calculated for a constant curvature
space-time background. We have demonstrated that the Hilbert-Einstein gravity
can be induced due to the quantum dynamics of torsion via formation
of a stable gravito-magnetic condensate.
One should notice that in our paper we have treated the metric as a fixed metric
of the classical space-time background while the contortion supposed to be
a quantum field. Such a treatment of metric is not merely satisfactory from the
conceptual point of view since one has to assume the pre-existence
of space-time with a metric given a priori.
One possible way to resolve that problem is to extend the Lorentz gauge group
to Poincare one, as it was mentioned in the Introduction.
In that case the gauge potential of Poincare group, the vielbein,
obtains dynamical content on equal footing with torsion.
Another interesting possibility is to consider Gauss-Bonnet-type gravity
model with torsion \cite{pakGB}. The model in the absence of torsion reduces
to a pure topological theory with arbitrary metric. Surprisingly,
within this model the torsion has the same number of physical degrees of freedom
for its spin two field component as the metric tensor.
This provides an additional argument supporting our conjecture that
torsion can play an important role as a quantum counterpart to the metric.
Possible implications of our results in cosmology of early Universe
will be considered elsewhere.

{\bf Acknowledgements}

Authors are grateful to Prof. Y.M. Cho for numerous discussions
and Prof. S.D. Odintsov for the interest to work.
Authors acknowledge Dr. M.L. Walker for the careful reading of the manuscript
and useful remarks.
One of the authors (DGP) thanks Prof. B.M. Zupnik for kind hospitality
during his visit to JINR, BLTP, Dubna.
The work is supported in part by the ABRL Program of
Korea Science and Engineering Foundation (R14-2003-012-01002-0)
and by Brain Pool Program (032S-1-8).

\end{document}